# Computational Study of Bouncing and Non-bouncing Droplets Impacting on Superhydrophobic Surfaces

Prathamesh G. Bange and Rajneesh Bhardwaj[*]

Department of Mechanical Engineering,

Indian Institute of Technology Bombay, Mumbai 400076, India

[*]Corresponding author (rajneesh.bhardwaj@iitb.ac.in)

Phone: +91 22 2576 7534, Fax: +91 22 2572 6875

*Abstract*

We numerically investigate bouncing and non-bouncing of droplets during isothermal impact on superhydrophobic surfaces. An in-house, experimentally-validated, finite-element method based computational model is employed to simulate the droplet impact dynamics and transient fluid flow within the droplet. The liquid-gas interface is tracked accurately in Lagrangian framework with dynamic wetting boundary condition at three-phase contact line. The interplay of kinetic, surface and gravitational energies is investigated via systematic variation of impact velocity and equilibrium contact angle. The numerical simulations demonstrate that the droplet bounces off the surface if the total droplet energy at the instance of maximum recoiling exceeds the initial surface and gravitational energy, otherwise not. The non-bouncing droplet is characterized by the oscillations on the free surface due to competition between the kinetic and surface energy. The droplet dimensions and shapes obtained at different times by the simulations are compared with the respective measurements available in the literature. Comparisons show good agreement of numerical data with measurements and the computational model is able to reconstruct the bouncing and non-bouncing of the droplet as seen in the measurements. The simulated internal flow helps to understand the impact dynamics as well as the interplay of the associated energies during the bouncing and non-bouncing. A regime map is proposed to predict the bouncing and non-bouncing on a superhydrophobic surface with an equilibrium contact angle of 155º, using data of 86 simulations and the measurements available in the literature. We discuss validity of the computational model for the wetting transition from Cassie to Wenzel state on micro- and nanostructured superhydrophobic surfaces. We demonstrate that the numerical simulation can serve as an important tool to quantify the internal flow, if the simulated droplet shapes match the respective measurements utilizing high-speed photography.

*Keywords*:

Droplet impact dynamics, Superhydrophobic surface, Computational model, Finite-element method, Bouncing, Non-bouncing, Dynamic wetting at contact line

## 1   Introduction

In the last decade, superhydrophobic surfaces have attracted significant attention due to their potential technical applications. These surfaces exhibit equilibrium contact angle ($\theta_{eq}$) greater than 150º and the impact dynamics of water droplets on such surfaces is central to understand and design self-cleaning processes [1], drag reduction techniques [2], pesticide spray coating on plants [3], and surface cooling via spray evaporative cooling [4] and pool boiling [5, 6]. The physics associated with the droplet impact on a solid surface in isothermal condition is an interplay of several coupled phenomena. The incompressible and laminar fluid flow within the droplet is highly transient. For example, impact of a 3 microliter isopropanol droplet with 0.37 m s$^{-1}$ velocity takes around 7 ms to spread on a heated fused silica surface [7]. The liquid-gas interface exhibits large deformation with Laplace stresses and the dynamic wetting at the three-phase contact line. The droplet fate depends on the interplay of several forces, namely, inertial, viscous, surface tension (capillary), wetting and gravity forces and it could result in spreading, bouncing or splashing depending on the impact conditions, surface wettability and roughness (see review by Yarin [8]). In a notable work by Schiaffino and Sonin [9], a regime map was proposed to predict the droplet impact dynamics on a solid surface. This map considered inertial, viscous and surface tension forces, however, ignored wetting forces at the contact line. In particular, the wetting forces are larger during the impact on the superhydrophobic surface and should be accounted for, while predicting the droplet fate.

The computational modeling of the droplet impact dynamics is challenging and involves large deformation of the liquid-gas interface with the dynamic wetting at the contact line. Most of the previous numerical studies can be classified in two categories on the basis of numerical treatment of the liquid-gas interface, namely, interface capturing in Eulerian framework and interface tracking in Lagrangian framework. In the former method, interface is represented implicitly by a scalar function while, it is represented explicitly by a time-varying mesh that follows the flow-field in the latter.

The Eulerian methods are further classified as volume-of-fluid [10] and level-set method [11]. In the former method, the liquid-gas interface is defined by a scalar function that is 0 and 1 in gas and liquid cells, respectively. The free surface is defined via the cells with intermediate values of the scalar function. Pasandideh-Fard et al. [12] employed volume-of-fluid method to simulate the droplet impact on flat and inclined surfaces, and compared numerical data with measurements. On the other hand, in the level-set method, the liquid-gas interface is defined using a signed normal distance function measured from the interface and is equal to zero at the interface. Liu et al. [13] used a sharp interface, ghost fluid method to simulate the droplet impact on surfaces of arbitrary shape. The level-set functions were used to represent fluid-fluid and fluid-solid interfaces [13]. Similarly, Caviezel et al. [14] simulated adherence, bouncing and splashing of the droplet using the level-set method on hydrophobic surface with equilibrium contact angle ranging from $90^{\circ}$ to $140^{\circ}$. In the Eulerian framework, lattice Boltzmann method can also be employed, in which discrete Boltzmann equation is solved on a lattice mesh. Very recently, Randive et al. [15] utilized two-phase lattice Boltzmann method to simulate interaction of the droplet with the superhydrophobic surface.

Utilizing Lagrangian approach, Poulikakos and co-workers [16 - 21] developed a finite-element method based flow solver with full Navier-Stokes equations for simulating the droplet impact dynamics on a solid surface. Later, Attinger and co-workers [4, 22 - 25] developed this solver for the impact and evaporation dynamics of a pure liquid droplet and a droplet laden with colloidal particles. Recently, Sprittles and Shikhmurzaev [26] used an arbitrary-Lagrangian, finite-element method based computational model to simulate the bouncing of the droplet on a hydrophobic surface with equilibrium contact angle of $130^{\circ}$. The advantage of Lagrangian method over Eulerian method is that the former allows precise tracking of the deforming liquid-gas surface. In present paper, we employ an experimentally-validated, finite-element method based flow solver in the Lagrangian framework. The solver was reported by Bhardwaj and Attinger [23] and is built upon the model of Wadvogel and Poulikakos [18]. In Ref. [23], the kinetic wetting model of Blake and de Connick [27] was implemented in the solver and it was validated with measurements for water droplets on a hydrophilic surface.

Most of the experimental studies utilized high-speed photography to record the droplet impact dynamics on superhydrophobic surfaces. For instance, Richard and Quéré [28] recorded several bouncing cycles of a 0.8 mm diameter water droplet on a superhydrphobic surface with static advancing contact angle of 170º. Clanet et al. [29] studied the bouncing of a 2.5 mm water droplet impact with 0.83 m s$^{-1}$ impact velocity on a superhydrophobic surface with equilibrium contact angle of 170º and showed that the maximum spreading $d_{max}$ scales as $d_0 We^{1/4}$, where $d_0$ is the initial droplet diameter and $We$ is Weber number defined as, $We = \rho v_0^2 d_0 \gamma^{-1}$, where $\rho$, $v_0$ and $\gamma$ are the density, impact velocity and surface tension of the droplet, respectively. Studies by Bartolo et al. [30], Jung and Bhushan [31] and Tsai et al. [32] reported the bouncing of 2 mm water droplets on superhydrophobic surface with equilibrium contact angle of around 155º ± 3º. Chen et al. [33] compared the droplet impact dynamics on an "artificial dual-scaled" superhydrophobic surface and a lotus leaf, with impact velocity ranging from 0.08 to 3 m s$^{-1}$ and equilibrium contact angle greater than 160º for both cases. The authors showed that the lowest Weber number for the bouncing droplet is 0.1 while splashing occurs at larger Weber numbers ($We > 41$) [33].

In the last decade, several experimental studies focused on the droplet impact dynamics on micro- and nanostructured superhydrophobic surfaces. The wetting on such surfaces can be categorized as Wenzel [34] and Cassie [35] state. The liquid fills the region between the micro - or nanopillars on the surface in the former while air traps below the droplet in the latter (Figure 1). The wetting transition from Cassie to Wenzel state occurs if the droplet overcomes "the energy barrier between the two states" [36]. Several studies [30, 37, 38] showed that the wetting transition from Cassie to Wenzel state can be achieved by varying impact velocity or equilibrium contact angle and consequently it affects the droplet fate i.e. the bouncing to non-bouncing. For instance, Bartolo et al. [30] reported that a 2 mm water droplet with impact velocity higher than 0.8 m s$^{-1}$ "sticks" on a micropatterned polydimethylsiloxane (PDMS) surface due to the wetting transition. Similarly, Wang et al. [37] showed that the droplet fate changes from the bouncing to non-bouncing due to decrease in equilibrium contact angle from 163º to 140º during the wetting transition on a superhydrophobic carbon nanotube arrays. In addition, the transition depends on the droplet volume and geometrical parameters (diameter and pitch) of micropillars, as shown by Lafuma and Quéré [39], and Jung and Bhushan [38], respectively.

Most of the previous numerical studies [4, 12, 13, 16, 20, 22, 23] focused on the impact dynamics on hydrophillic surfaces ($\theta_{eq} < 90°$). A few numerical studies considered droplets on the hydrophobic surfaces ($90° < \theta_{eq} < 150°$). For example, Caviezel et al. [14] reported the droplet impact dynamics on the superhydrophobic surface and Sikarwar et al. [40] analyzed three-dimensional flow fields of pendant droplets on a hydrophobic surface. However, these studies ignored the dynamic wetting and assumed dynamic contact angle as constant at the contact line [14, 40]. The second issue in this arena is that the quantitative measurement of the internal flow is not trivial. Quantifying the internal flow may help to understand and design technical applications during the droplet deposition. In this context, Sakai et al. [41] measured internal flow using particle image velocimetry (PIV) in a sliding droplet on a hydrophobic surface. However, suspended tracer particles in the PIV technique may modify wetting behavior at the contact line and thereby impact dynamics. To the best of our knowledge, there is no experimental study reported thus far that quantify the internal flow for the bouncing or non-bouncing droplet on the superhydrophobic surface.

The objective of the present paper is twofold: first, to numerically investigate the effect of the impact velocity and equilibrium contact angle on the impact dynamics of the bouncing and non-bouncing droplets on a superhydrophobic surface in isothermal condition. Specifically, the present study focuses on the interplay of the kinetic, surface and gravitational energies during the droplet impact. We employ a state-of-the-art, in-house, experimentally-validated, finite-element method based computational model with the dynamic wetting at the three-phase contact line. The second objective is to match the simulated droplet shapes with high-speed photography measurements in order to predict the internal flow using the numerical simulation with reasonable fidelity. The information on the internal flow further helps to corroborate the understanding of the interplay of the associated energies in the first objective.

The paper is organized as follows. In section 2, we present brief details of the computational model. Second, we investigate the effect of the impact velocity (section 3.1), and the equilibrium contact angle (section 3.2). Third, we propose a regime map for the bouncing and non-bouncing droplets on the superhydrophobic surface with an equilibrium contact angle of 155º (section 3.3). Finally, we present limitations of the computational model and directions for the future work (section 3.4).

## 2  Computational model

In this section, we present the computational model to simulate the droplet impact dynamics on the superhydrophobic surface in isothermal condition. The model employed in the present study was developed initially by Poulikakos and co-workers [16 - 21] and later by Attinger and co-workers [4, 22 - 25]. Specifically in this paper, we employed the model reported by Bhardwaj and Attinger [23], that was built upon model of Wadvogel and Poulikakos [18]. It solves radial and axial components of Navier-Stokes equations (eq. 1) in axisymmetric, cylindrical coordinates for laminar flow within the droplet. An artificial compressibility method is employed to transform the continuity equation (eq. 2) into a pressure evolution equation [42] and a small Mach number $M$ = 0.001 is used in the model. We neglect the effect of convection in surrounding air, deformation of the liquid-gas interface due to air overpressure developed underneath the droplet just before the impact and bubble entrapped in the droplet just after the impact. An analytical model described in Appendix 6.1 considers lubrication approximation for the squeezing air flow in thin film [43] and the deformation of the liquid-gas interface due to the air overpressure [44, 45]. Using this model, we calculate the maximum size of the bubble [44] that would enter into the droplet by equating the droplet internal pressure and air overpressure in the film. The maximum volume of the bubble entering into the droplet for the cases in the present study is six orders of magnitude lesser than the droplet volume and thereby, it is unlikely to influence the internal fluid flow and impact dynamics. In addition, the maximum deformation of the liquid-gas interface is three orders of magnitude lesser than the droplet diameter and can be neglected. Hence, neglecting the squeezing air film in the computational model can be justified. However, the thin air film underneath the droplet is important at larger impact velocity [46] or during the impact on perfectly hydrophilic surface with almost negligible equilibrium contact angle [47, 48]. Thus, the modeling of the air film squeezing by the droplet and associated bubble formation should be accounted for in such cases. The evaporation due to liquid-vapor diffusion in the ambient is also neglected and justification for this assumption was given by Bhardwaj and Attinger [23]. In the following sub-sections, we briefly describe the governing equations, boundary conditions and numerical scheme with code validation..

## 2.1 Governing equations

The dimensionless governing equations are given below and detailed derivation can be found in previous papers [18, 22, 23]. The length, velocity and time scale used for non-dimensionalization are $d_0$, $v_0$ and $d_0/v_0$, respectively, where $d_0$ and $v_0$ are the initial droplet diameter and impact velocity, respectively. The momentum and mass conservation are written in vector form as follows,

$$\frac{D\mathbf{V}}{D\tau} - \nabla \cdot \mathbf{T} + \frac{\mathbf{n_z}}{Fr} = 0, \tag{1}$$

$$\frac{DP}{D\tau} + \frac{\nabla \cdot \mathbf{V}}{M^2} = 0, \tag{2}$$

where $\mathbf{V}$ is dimensionless velocity vector $\mathbf{V} = (U, V)$; $Fr$ is Froude number, $Fr = v_0^2 d_0^{-1} g^{-1}$, where $g$ is gravitational acceleration (9.81 m s$^{-2}$); $P$ is dimensionless pressure, $P = p\rho^{-1}v_0^{-2}$, where $p$ and $\rho$ are dimensional pressure and droplet density, respectively; $\tau$ is dimensionless time; and $M$ is Mach number, $M = v_0 c^{-1}$, where $c$ is speed of sound. The unit vector $\mathbf{n} = \mathbf{n_r} + \mathbf{n_z}$ is normal to the liquid-gas interface, $\mathbf{n_r}$ and $\mathbf{n_z}$ are the radial and axial components of the vector $\mathbf{n}$, respectively, and $\mathbf{T}$ is the stress tensor, defined as follows:

$$\mathbf{T} = \begin{bmatrix} \overline{\sigma_{RR}} & \overline{\sigma_{RZ}} & 0 \\ \overline{\sigma_{ZR}} & \overline{\sigma_{ZZ}} & 0 \\ 0 & 0 & \overline{\sigma_{\theta\theta}} \end{bmatrix}, \tag{3}$$

The dimensionless stress tensor terms $\bar{\sigma}_{ij}$ are expressed as:

$$\bar{\sigma}_{RR} = -P + \frac{2}{Re}\frac{\partial U}{\partial R}, \bar{\sigma}_{\theta\theta} = -P + \frac{2}{Re}\frac{U}{R},$$
$$\bar{\sigma}_{RZ} = \bar{\sigma}_{ZR} = \frac{1}{Re}\left(\frac{\partial U}{\partial Z} + \frac{\partial V}{\partial R}\right), \bar{\sigma}_{ZZ} = -P + \frac{2}{Re}\frac{\partial V}{\partial Z}, \tag{4}$$

where $Re$ is Reynolds number, defined as $Re = \rho v_0 d_0 \mu^{-1}$, where $\mu$ is the dynamic viscosity.

### 2.1.1 Initial and boundary conditions

The initial and boundary conditions are summarized in Figure 2. A spherical droplet is considered at time $\tau = 0$:

$$\tau = 0: U = 0, V = -1, \tag{5}$$

The boundary condition along the liquid-gas interface is balance of the Laplace pressure and viscous stresses [18].

$$(\mathbf{T})^T \cdot \mathbf{n} = -2\frac{\bar{H}}{We}\mathbf{n}, \tag{6}$$

In the above equation, $\bar{H}$ is dimensionless free surface curvature. Marangoni stresses are neglected due to absence of thermal gradient along the liquid-gas interface and shear stress at the interface is also neglected ($\mu_{droplet} \gg \mu_{air}$). The axisymmetric boundary condition is applied along the z-axis:

$$\text{At } R = 0: U = 0, \frac{\partial V}{\partial R} = 0, \tag{7}$$

No-penetration boundary condition $V = 0$ is applied along r-axis ($Z = 0$).

### 2.1.2 Boundary conditions at the contact line

It is well-documented in several reviews [49-55] that the no-slip boundary condition at the contact line generates a non-integrable shear stress singularity at the contact line. As reviewed by Ren and E. [54], previous methods which addressed the singularity can be classified in two categories, namely, diffuse interface models (see representative papers of Shikhmurzaev and co-workers [26, 52, 56]) and slip models [51, 53, 55]. In the former, the kinematic boundary condition at the liquid-gas interface is modified to allow nonzero mass flux across the interface near the contact line [54]. In the latter, a small finite region of the liquid at the contact line is assumed to slide along the substrate and Navier-slip condition is applied near the contact line [54, 55]. In the present work, the slip is only allowed for the contact line and the slip-region is not completely numerically resolved due to the following reasons. The slip region near the contact line is on the order of nanometers [55, 57] and computational scale in the bulk region is on the order of millimeters. In order to numerically resolve the flow in the slip region, an ultrafine grid is required as compared to that in the bulk of the droplet. We do not completely resolve the flow in the slip region for the sake of computational tractability. However, we make sure that the grid-size in the slip region is fine enough to obtain the grid-size independent results for the flow-field inside the droplet (see section 2.1.3). In this context, same flow-field in the bulk region were obtained using different slip models in the slip region by Qian et al. [58], which demonstrates that the flow-field in the slip region does not alter the flow-field in the bulk, consistent with the conclusions by Dussan [50].

In addition to the elimination of the shear stress singularity at the contact line, a description of the contact line velocity with the dynamic contact angle is needed to enforce the boundary condition at the contact line (eq. 6). The contact line velocity may be prescribed in terms of the dynamic contact angle via hydrodynamic or kinetic wetting approach (see discussion in section 1.1 in Bhardwaj and Attinger [23]). In the present study, we prescribe the dynamic contact line velocity in terms of the dynamic contact angle using the kinetic model of Blake and and de Connink [27]. We apply two essential (velocity) conditions at the contact line as recommended by Bach and Hassager [59]. These two boundary conditions are, namely, the dynamic contact line velocity in the radial direction ($u = u_{CL}$) and no-penetration boundary condition ($v = 0$) in the axial direction. The no-slip boundary condition is applied along $r$-axis except at the contact line ($R_{wetted}$, 0), at which the dynamic contact line velocity boundary condition is applied,

$$\text{At } R \neq R_{wetted}, U = 0,$$
$$\text{At } R = R_{wetted}, U = \frac{u_{CL}}{v_0}, \quad (8)$$

where $u_{CL}$ is dynamic velocity at the contact line (m s$^{-1}$), given by the kinetic wetting model of Blake and de Connink [27], that describes the wetting as a dynamic adsorption/desorption process of liquid molecules on the surface. The wetting boundary condition in the finite-element flow solver was implemented by Bhardwaj and Attinger [23] and here we briefly present the wetting model. The contact line velocity, $u_{CL}$, during the wetting of liquid on a solid surface is expressed as [27]:

$$u_{CL} = \frac{2\kappa_s^0 h\lambda}{\mu v_L} \sinh\left[\frac{\gamma}{2nk_BT}(\cos\theta_{eq} - \cos\theta)\right] = \frac{2K_w\lambda}{\mu} \sinh\left[\frac{\gamma}{2nk_BT}(\cos\theta_{eq} - \cos\theta)\right], \quad (9)$$

where $\theta_{eq}$ and $\theta$ are the equilibrium and dynamic contact angle, respectively. The equilibrium contact angle $\theta_{eq}$ is taken from the measurements reported in the literature and dynamic contact angle $\theta$ is obtained from the numerical simulation. In above equation, $\gamma$ is surface tension [N m$^{-1}$], $\mu$ is viscosity [Pa s] and $v_L$ is molecular volume of liquid [m$^3$], $\lambda$ is molecular displacement length (O(angstrom)), $n$ is the number of adsorption sites per unit area, $k_B$ is the Boltzmann's constant ($1.3806503 \times 10^{-23}$ m$^2$ kg s$^{-2}$ K$^{-1}$), $h$ is the Planck's constant ($6.626 \times 10^{-34}$ m$^2$ kg/s), $T$ is the absolute temperature ($T = 298$ K) and $\kappa_s^0$ is a constant involving the molecular equilibrium displacement frequency (s$^{-1}$), $\kappa_w^0$ [27]:

$$\kappa_w^0 = \kappa_s^0 \left( \frac{h}{\mu v_L} \right), \qquad (10)$$

The wetting parameter ($K_w = \kappa_s^0 h/v_L$) is used as a parameter controlling the wetting speed and in the previous studies [4, 23], it was obtained by matching the droplet shapes obtained by the numerical simulation and high-speed visualization, assuming that $\lambda = 2 \times 10^{-10}$ m and $n = \lambda^{-2}$ [20, 60]. In the present work, we follow the same strategy to obtain $K_w$ and it is taken as $1 \times 10^6$ Pa in all simulations presented in this paper.

### 2.1.3 Numerical scheme and code validation

An implicit method proposed by Bach and Hassager [59] is utilized for the numerical integration of the fluid dynamics equations in time and details of the algorithm can be found in Ref. [18]. The computational domain is discretized using mesh of triangular elements and the computational model is solved using Galerkin finite-element method. Linear shape functions are used for velocity as well as pressure. The computational model was extensively validated for the impact dynamics for the droplets of molten metal [18], water [23] and isopropanol [4] in previous papers. The details of grid-size and time-step independence study can be found in Ref. [23]. The thermophysical properties are considered for water at ambient temperature (25°C) [61]: density $\rho = 1000$ kg m$^{-3}$, dynamic viscosity $\mu = 9.0 \times 10^{-4}$ Pa s and surface tension $\gamma = 7.2 \times 10^{-2}$ N m$^{-1}$.

## 3   Results and Discussions

In this section, we investigate the droplet impact dynamics on the superhydrophobic surface using the computational model described in section 2. The numerical simulations are compared with the available measurements and analytical models, wherever possible. In all simulations, the Reynolds and Weber number are selected such that splashing does not occur ($Re < 1200$, $We < 7$) [33]. This section is organized as follows:

- First, we examine effect of the impact velocity on the bouncing and non-bouncing in section 3.1. We vary the impact velocity keeping the equilibrium contact angle and droplet diameter constant in cases 1 and 2, listed in Table 1. The parameters of the cases 1 and 2 correspond to the measurements by Tsai et al. [32] for the bouncing and non-bouncing droplets, respectively.

- Second, we investigate effect of the equilibrium contact angle on the bouncing and non-bouncing in section 3.2. We vary the equilibrium contact angle keeping the impact velocity and droplet diameter constant in cases 4 and 5, listed in Table 1. The parameters of the two cases correspond to the measurements by Jung and Bhushan [31] for the bouncing and non-bouncing droplets, respectively.

- Finally, a regime map is presented in section 3.3 to predict the bouncing and non-bouncing droplets using data of 86 simulations and the measurements reported in the literature [30 - 32].

While examining the effect of the impact velocity and equilibrium contact angle in sections 3.1 and 3.2, the droplet diameter is considered smaller than capillary length for water ($l_c = \sqrt{\gamma/(\rho g)}$ = 2.7 mm) so that gravity does not influence the impact dynamics. The cases considered in the presented study exhibit smaller viscous dissipation inside the droplet and it will be important at lower Reynolds number or larger viscosity. The energy associated with the viscous dissipation and initial kinetic energy of the droplet scale as [29] $\mu v_0 d_{max}^3 / h$, and $\rho d_0^3 v_0^2$, respectively, where $\mu$ is dynamic viscosity, $v_0$ is impact velocity, and $d_{max}$ and $h$ are the maximum droplet width and height, respectively, at the instance of the maximum spreading. Using scaling analysis [29], the energy associated with the viscous dissipation is estimated lesser than 12% of the initial kinetic energy for all cases considered in the present study.

## 3.1 Effect of impact velocity

In order to investigate the effect of the impact velocity, we performed numerical simulations for the two cases of impact velocity, keeping equilibrium contact angle and droplet diameter as constant. The impact velocity ($v_0$) are 0.29 m s$^{-1}$ and 0.13 m s$^{-1}$ in the two cases (Table 1, case 1 and 2, respectively), and the droplet diameter and equilibrium contact angle are 2 mm and 155º, respectively. Figure 3 presents simulated droplet shapes, contours of velocity magnitude and instantaneous streamlines for case 1 ($v_0$ = 0.29 m s$^{-1}$). Computer animation for this case is provided as supplementary data [62]. The initial spreading is driven by the large kinetic energy ($t$ < 3.43 ms) as illustrated by radially outward streamlines with large deformation on the free surface. At instance of maximum spreading ($t \sim 3.43$ ms), the droplet height reaches its minimum value and the maximum width $D_{max}$ (illustrated in Figure 2B) is 1.3. According to analytical model by Clanet et al. [29], the maximum dimensionless droplet width, $D_{max}$, scales as follows:

$$D_{max} \sim We^{1/4}, \qquad (11)$$

where $We$ is Weber number, with $We > 1$ as condition of the validity of the model. The value obtained by the analytical model is $D_{max}$ = 1.25, consistent with the simulated value ($D_{max}$ = 1.3). The droplet recoils at $t$ = 4.11 ms due to conversion of the surface energy to the kinetic energy and the internal flow reverses to radially inward. The flow becomes axially upward from $t$ = 4.8 ms to 10.96 ms due to further increase in the kinetic energy. The flow reverses again at $t$ = 11.64 ms to axially inward due to competition between the surface and kinetic energy. Lastly, the droplet bounces off the surface at $t$ = 13.7 ms. The liquid-gas interface breakup cannot be handled by the present model and thus we stop the simulation at incipience of the bouncing.

Figure 4 describes numerical simulation for case 2 ($v_0$ = 0.13 m s$^{-1}$) and computer animation is provided as supplementary data [63]. As seen earlier in case 1, the droplet spreads initially due to larger kinetic energy until $t$ = 4.15 ms and it recoils at $t$ = 5.54 ms due to the conversion of the surface energy to the kinetic energy with the internal flow reversal, shown by the simulated streamlines. Subsequently, the free surface oscillates with the flow reversal from axially upward to axially downward, due to competition between surface and kinetic energy and thereby droplet does not bounce.

Figure 5(a) plots comparisons of dimensionless droplet maximum width ($D_{max}$) and height ($H_{max}$) between the bouncing (case 1) and non-bouncing (case 2) droplets. The droplet maximum width and height are illustrated in Figure 2B. Due to larger Weber number, the maximum spreading and height in case 1 are 12% larger and 25% lesser than that in case 2, respectively. Similarly, the time taken for the maximum spreading in case 1 is 53% shorter than that in case 2. As discussed earlier, the droplet bounces in case 1 while it exhibits the non-bouncing in case 2 with oscillations on the free surface.

The bouncing and non-bouncing can be explained using the interplay of the energies during the impact as follows. The time-varying kinetic ($E_k$), surface ($E_s$) and gravitational ($E_g$) energies are plotted for the bouncing (case 1) and non-bouncing (case 2) cases in Figure 5(b) and Figure 5(c), respectively. In addition, we plot the sum of all energies at any instance ($E = E_k + E_s + E_g$) and sum of the initial surface and gravitational energy ($E_{s0} + E_{g0}$) for both cases. The mathematical expressions of the energies are given in Appendix 6.2 and are taken from Refs. [64, 65]. As plotted in Figure 5(b) for the bouncing case (case 1), the gravitational energy is lesser than 10% of the total energy at all times. The surface energy ($E_s$) starts increasing at the expense of kinetic energy ($E_k$) after the droplet impact and at the instance of maximum spreading ($t \sim 3.9$ ms), the surface and kinetic energy reaches to the maximum and minimum, respectively. After the recoiling, the former decreases at the expense of the latter until 6 ms. According to Mao et al. [64], if the total energy, $E = E_k + E_s + E_g$, at the instance of the maximum recoil (i.e. when the droplet height reaches to the maximum after the recoiling) exceeds the initial surface and gravitational energy, $E_{s0} + E_{g0}$, the droplet bounces off the surface, otherwise not. In other words, the droplet bounces if it has sufficient energy to recover its original spherical shape and diameter. Note that Mao et al. [64] neglected the gravitational energy in their analysis. In case 1, the total energy at the instance of the maximum recoil exceeds the initial surface and gravitational energy ($E > E_{s0} + E_{g0}$) in Figure 5(b) and the droplet bounces off the surface. The energies for case 2 are plotted in Figure 5(c). The initial kinetic energy is around one order of magnitude lesser the initial surface energy because of larger Weber number ($We = 2.34$) as compared to that in case 1. The gravitational energy is on the same order as the initial kinetic energy and does not vary much during the impact. The conversion of the surface and kinetic energy into each other is similar to that in case 1. However, at the instance of maximum recoiling, the total energy at the instance of the maximum recoil is on the order of the initial surface and gravitational energy ($E \sim E_{s0} + E_{g0}$) and thereby, droplet does not bounce.

The bouncing time and time-period of the oscillation of the free surface in case 1 and case 2, respectively, scale as [9, 23, 29, 66]

$$t \sim \sqrt{\frac{\rho d_0^3}{\gamma}}, \qquad (12)$$

where $\gamma$, $\rho$ and $d_0$ are surface tension, density and initial droplet diameter, respectively. The bouncing time and time-period of the oscillation obtained by the respective simulations are 13 and 10 ms, respectively, that are on the same order as estimated by the analytical model (10.5 ms, eq. 12).

### 3.1.1 Comparison with experiments by Tsai et al. [32]

The numerical results of the bouncing (case 1) and non-bouncing (case 2) droplets described in previous section are compared with high-speed photography measurements reported by Tsai et al. [32] in Figure 6 and Figure 7, respectively. The results are compared in terms of the time-varying droplet maximum width and height. In the insets of Figure 6 and Figure 7, the recorded droplet shapes (left) are also compared with the respective simulated ones with instantaneous streamlines (right), at different time instances. The numerical simulation predicts the droplet spreading, recoiling and the bouncing as seen in measurements for case 1 with reasonable accuracy (Figure 6). The insets show good agreement of the droplet shapes at different time instances and surface level is shown by a horizontal line in the images obtained by the simulation (right). The maximum errors in the calculated maximum width and height as compared to the respective measurements are around 8% and 12%, respectively, at $t = 8$ ms. At the time of the bouncing, the maximum width obtained in simulations is same as in measurements and error in the height is around 5%. The bouncing time obtained from the simulation is 8% larger than that in the experiments.

Figure 7 compares the time-varying maximum width and height obtained in the simulation and measurement [32] for the non-bouncing droplet (case 2). The time-variations of the calculated maximum width and height show the initial spreading of the droplet, recoiling and subsequent free surface oscillations, as seen in the measurement by Tsai et al. [32]. The free surface oscillates due to the competition between the kinetic and surface energy. The agreements for the dynamic behavior of the maximum width and height, and droplet shapes are good. After the droplet becomes sessile, the calculated values of maximum width and height are 2% and 7% larger as compared to the respective experimental ones. We note slightly large errors of +14% and -11% in the maximum width at 0.33 and 5.7 ms, respectively, and -22% and +19% in maximum height at 3 ms and 5.7 ms, respectively. This may be attributed to uncertainty in initial droplet diameter (2 ± 0.1 mm) and equilibrium contact angle (155º ± 3º) in measurements by Tsai et al. [32]. Overall, the simulations capture several features recorded by the high-speed visualization and confirm the bouncing and non-bouncing of the droplet as seen in experiments [32] in Figure 6 and Figure 7, respectively.

### 3.1.2   Comparison with experiments by Wang et al. [37]

In this section, we compare numerical data with measurement of Wang et al. [37] for 2 mm bouncing water droplet on a superhydrophobic surface with equilibrium contact angle of 163º (listed as case 3 in Table 1). Figure 8 shows the comparison between the simulated and measured wetted diameter with a good agreement. The simulation correctly predicts the droplet bouncing and captures well the time-variation of the wetted diameter as recorded in the measurements of Wang et al. [37]. The error in the calculated values with respect to the experiments are around 10%. The bouncing times obtained by the simulation and measurement are in very good agreement (14 ms).

## 3.2   Effect of equilibrium contact angle

In order to investigate the effect of equilibrium contact angle, we describe numerical simulations for two cases of equilibrium contact angle $\theta_{eq}$, keeping impact velocity and diameter as constant. The equilibrium contact angles are 158º and 91º in the two cases considered (Table 1, case 4 and 5, respectively), and the droplet diameter and impact velocity are 2 mm and 0.44 m s$^{-1}$, respectively. Figure 9 illustrates contours of velocity magnitude, simulated instantaneous streamlines and evolution of the droplet shape for case 4 ($\theta_{eq}$ = 158º). Computer animation for this case is provided as supplementary data [67]. The initial spreading is mostly due to the kinetic energy ($t$ < 3.45 ms) and the droplet takes a toroidal shape at $t$ = 4.18 ms. The maximum droplet width $D_{max}$ = 1.52 at $t$ = 3.45 ms is consistent with the one obtained by analytical model (1.44, eq. 11). After the maximum spreading, the droplet recoils due to the conversion of surface energy into the kinetic energy and the flow reversal occurs at the incipience of recoiling at $t =$ 4.18 ms. The streamlines illustrate the axially upward flow pattern from 4.86 to 13.9 ms as the droplet recoils and bounces off the surface at 13.9 ms due to larger total energy as compared to initial surface energy.

Figure 10 shows the impact of the droplet for case 5 ($\theta_{eq} = 91°$) and computer animation is provided as supplementary data [68]. We note similar features in case 5 as described in section 3.1 for case 2. Initially, the droplet spreads under the influence of the kinetic energy and after recoiling, the total kinetic and surface energy at the maximum recoiling is lesser than the initial surface energy. Thus, the droplet does not bounce and the free surface oscillates due to competition between the kinetic and surface energy.

Figure 11(a) compares the time-varying maximum width and height for the cases 4 and 5. The maximum spreading is almost same in the two cases because the kinetic energy in the two cases are on the same order, however, the droplet retracts more during recoiling in case 4 because of lower surface wettability. The time taken for the maximum spreading in the two cases is almost same. As discussed earlier, the droplet bounces in case 4, while it does not bounce in case 5 and the free surface oscillates with a certain frequency.

We explain the bouncing and non-bouncing using the interplay of the associated energies as follows. We plot the evolution of kinetic energy ($E_k$), surface energy ($E_s$) and gravitational energy ($E_g$), total energy ($E$) and the sum of the initial surface and gravitational energy ($E_{s0} + E_{g0}$) for the bouncing and non-bouncing cases in Figure 11(b) and Figure 11(c), respectively. The gravitational energy is less than 10% of the total energy and does not vary much during the impact in both cases. We observe similar trends for kinetic and surface energies in case 4 similar to case 1 as discussed in section 3.1. The surface energy increases at the expense of kinetic energy until the maximum spreading and after the recoiling, the kinetic energy increases at the expense of surface energy. Due to larger surface wettability in case 5 ($\theta_{eq} = 91°$) as compared to case 4 ($\theta_{eq} = 158°$), the surface energy decreases much faster after the recoiling in case 5, as shown in Figure 11(c). At the instance of the maximum recoiling, the droplet bounces in case 4 because the total energy exceeds the initial surface and gravitational energy, $E > E_{s0} + E_{g0}$, while it does not bounce in case 5 because the total energy is less than the initial surface and gravitational energy, $E < E_{s0} + E_{g0}$.

The calculated bouncing time in case 4 (13.7 ms) is on the same order as estimated by the analytical model (10.5 ms, eq. 12). In addition, the simulated bouncing time in this case ($v_0 = 0.44$ m s$^{-1}$) is almost same as calculated in case 1 in section 3.1 (13 ms, $v_0 = 0.29$ m s$^{-1}$). Thus, the simulated bouncing times are consistent with finding by Richard et al. [66] that the bouncing time does not depend on the impact velocity in range [0.2 - 2.3 m s$^{-1}$]. The calculated time-period of the oscillation $t_{osc} \sim 10$ ms in case 5 is close to the value obtained by the analytical model (8.3 ms, eq. 12).

### 3.2.1 Comparison with experiments by Jung and Bhushan [31]

Figure 12 and Figure 13 compares the simulation results for cases 4 and 5 listed in Table 1 and described in section 3.2 with high-speed visualization measurements by Jung and Bhushan [31], respectively. The simulated and recorded droplet shapes are compared in the insets. In the measurements by Jung and Bhushan [31], the initial position of the droplet was recorded at a certain distance from the surface and recorded times at different instances were relative to the initial position. In the simulations, we considered the initial position when the droplet touches the surface ($t = 0$ ms). In order to be consistent with the numerical model, we modified the recorded times accordingly in Figure 12 and Figure 13. Nonetheless, we compared the initial droplet shapes obtained by the simulation and measurement in order to compare the droplet shape, as shown in top-left insets in Figure 12 and Figure 13. The calculated and measured time-varying droplet maximum width and height in Figure 12 are in good agreement. The error in the simulated maximum width and height as compared to the recorded values at instance of the bouncing (~11.83 ms) are 3% and 8%, respectively. The droplet shapes predicted by the numerical simulations are confirmed by the high-speed visualization and computational model is able to reconstruct the bouncing of the droplet, as recorded in the experiments [31]. The difference in the droplet shape at $t = 3.8$ ms can be attributed to the fact that the image obtained by the simulation is a radial cut while measurement is a side image.

Figure 13 shows the comparison between the time-varying droplet maximum width and height obtained by the simulation and measurements [31] for case 5. The droplet spreading, recoiling and free surface oscillations recorded in experiments are reproduced by the simulation. The droplet dynamics and shapes are in good agreement and the numerical simulation correctly predicts the non-bouncing of the droplet, After the droplet becomes sessile ($t \sim 25.6$ ms), the error in the simulated droplet maximum width and height as compared to the recorded values are around 1% and 5%, respectively. In general, the simulations reconstruct the bouncing and non-bouncing with reasonable fidelity as seen in experiments [31], in Figure 12 and Figure 13, respectively.

The contact angle hysteresis ($\theta_{adv}$ - $\theta_{rec}$) plays an important role in determining the bouncing, where $\theta_{adv}$ and $\theta_{rec}$ are advancing and receding contact angle, respectively [28, 31, 33, 37, 69]. The droplet bounces off during recoiling if droplet kinetic energy ($\sim \rho d_0^3 v^2$) overcomes surface energy ($\sim \gamma d_0^2 |\cos\theta_{adv} - \cos\theta_{rec}|$) [33, 69]. The expression of the critical bouncing threshold velocity ($v_c$) thus obtained is as follows [33, 69],

$$v_c \sim \left( \frac{\gamma |\cos\theta_{adv} - \cos\theta_{rec}|}{\rho d_0} \right)^{1/2} \quad (13)$$

Jung and Bhushan [31] reported the measurements of the contact angle hysteresis as 12° and 87°, for case 4 (bouncing) and case 5 (non-bouncing) listed in Table 1. Using the reported values of the advancing and receding contact angles in Ref [31] in eq. 13, the estimated threshold velocities for the bouncing and non-bouncing cases are 0.08 and 0.22 m s$^{-1}$, respectively. Thus, lower contact angle hysteresis helps in the bouncing since the droplet requires lower threshold velocity.

### 3.3 Regime map to predict the bouncing and non-bouncing droplets

We propose a regime map to predict the bouncing and non-bouncing droplets on a superhydrophobic surface. In this context, a regime map based on numerical simulations was proposed by Caviezel et al. [14] for the non-bouncing, bouncing and splashing droplets on a hydrophobic surface. In this study [14], the range of Weber numbers and equilibrium contact angle were taken as [0.1, 100] and [90°, 140°], respectively, keeping Reynolds number as constant ($Re$ = 800). In present study, the proposed regime map plots data of two simulations in Table 1 (cases 1 and 2), 84 simulations listed in Table 2, and available measurements at $\theta_{eq}$ = 155° ± 3° [30 - 32, 37]. In the numerical simulations, we varied Reynolds and Weber number in range [100, 1200] and [1, 7], respectively, keeping equilibrium contact angle constant ($\theta_{eq}$ = 155°). The corresponding range of the droplet diameter and impact velocity are [1.6 × 10$^{-2}$, 16] mm and [0.067, 5.6] m s$^{-1}$, respectively (Table 2). In Figure 14, the bouncing and non-bouncing regimes are shown with filled and hollow symbols, respectively and demarcated by a dashed line. The bouncing regime is characterized by larger Reynolds and Weber number ($Re > 800$ and $We > 3$) while the non-bouncing corresponds to smaller Reynolds and Weber number ($Re < 500$ and $We < 2$). Threshold values of Reynolds and Weber number exist for the transition from the non-bouncing to bouncing regime. The measurements reported in the literature [30 - 32] are also plotted in Figure 14 along with the numerical data. In general, measurements points are consistent with the bouncing and non-bouncing regimes as suggested by the numerical data, with exception of one measurement by Tsai et al [32]. The source of this discrepancy is not clear to us and it may be attributed to different wetting characteristics of carbon nanofibers substrate in the experiments of Tsai et al [32].

### 3.4 Limitations of the computational model

The present computational model is axisymmetric and valid only for orthogonal impact of the droplet on the substrate. However, in case of non-orthogonal impact, three-dimensional effects are important, as shown by Pasandideh-Fard et al. [12] and should be accounted for in the model. The second major limitation of the model is that it does not account for the Cassie to Wenzel wetting transition during the droplet impact on micro- or nanostructured hydrophobic or superhydrophobic surface. As discussed in section 1, the droplet fate on such surfaces changes from the bouncing to non-bouncing due to the wetting transition from the Cassie to Wenzel state [30, 37, 38]. We examine whether the present computational modeling is adequate enough to simulate the non-bouncing due to the transition. Wang et al. [37] showed that a 2 mm water droplet does not bounce on superhydrophobic carbon nanotube arrays due to the transition, which reduces the equilibrium contact angle from 163º to 140º. We perform a numerical simulation using our model corresponding to the measurement reported by Wang et al. [37], in which the transition occurs. The equilibrium contact angle is taken as 140° in the model (case 6 in Table 1). The simulation predicts the bouncing (not shown) while the measurement in case 6 recorded the non-bouncing. The discrepancy between the numerical and experimental data suggests that the computational modeling presented in this work is not suitable for the case with the wetting transition. In order to simulate such case, the present modeling should be extended to account for the transition. In this context, Lattice Boltzmann method was recently employed for simulating the droplet impact [70] and evaporation [71] on the micropillars. Finally, at larger capillary numbers, very fine mesh elements near the contact line are needed to numerically resolve significant bending of the liquid-gas interface near the contact line [56, 72]. The present modeling does not consider it and such cases will be considered in future.

## 4 Conclusions

A numerical investigation of the droplet impact dynamics on the superhydrophobic surface is presented using an in-house, experimentally validated, finite-element method based computational model [23]. We systematically varied the impact velocity and equilibrium contact angle to investigate the interplay of the kinetic, surface and gravitational energies during the bouncing and non-bouncing of the droplet. The numerical simulations show that the droplet bounces off the surface if the total energy of the droplet at the instance of maximum recoiling exceeds initial surface and gravitational energy, otherwise not. In case of the non-bouncing droplet, the free surface oscillates due to competition between the kinetic and surface energy. The calculated maximum width, bouncing time and time-period of the oscillation are consistent with the analytical models available in the literature. The computational model is able to reconstruct the bouncing and non-bouncing of the droplets, as recorded in the reported measurements. The calculated maximum droplet width and height with respect to the measurements are reasonably accurate and verify the numerical simulations. The comparisons of the time-varying droplet shapes are also in good agreement and qualitatively confirm the simulated flow fields. A regime map is proposed to predict the bouncing and non-bouncing droplets, based on data of 86 simulations and available measurements. In this map, the range of Reynolds and Weber number are [100, 1200] and [1, 7], respectively, keeping equilibrium contact angle constant $\theta_{eq} = 155°$. In general, the bouncing and non-bouncing regimes suggested by the numerical simulations are consistent with the available measurements. Finally, we examined the validity of the present model to predict the bouncing and non-bouncing during the wetting transition from Cassie to Wenzel state for micro- and nanostructured superhydrophobic surfaces. We concluded that the present computational model cannot be used if the wetting transition occurs during the impact. Since measuring the internal flow during the droplet impact is not trivial, the present study demonstrates that the numerical simulation can serve as an important tool to quantify the fluid flow, after matching of the simulated droplet shapes with the measurements. The characterization of the internal flow could help in designing technical applications such as self-cleaning processes [1] and drag reduction techniques [2].

## 5 Acknowledgements

We gratefully acknowledge financial support from Department of Science and Technology (DST), New Delhi, through fast-track scheme for young scientists. This work was also partially supported by an internal grant from Industrial Research and Consultancy Center (IRCC), IIT Bombay.

## 6 Appendix

### 6.1 Analytical model for estimating size of entrapped bubble during the droplet impact

In order to estimate the size of the bubble that enters in the droplet during the impact, we consider the lubrication flow in squeezed air film between the droplet and substrate. The effect of the squeezed air film becomes important just before the instance of impact and we consider a droplet of diameter $d_0$ approaching to the substrate with impact velocity $v_0$, as illustrated in Figure 15(a). The air overpressure in the film deforms the liquid-gas interface near the impact location, as shown in inset of Figure 15(b). As described by Leal [43], we consider the lubrication equation for air overpressure in the squeezed air film in axisymmetric geometry,

$$12\mu_{air}\frac{\partial h}{\partial t}=\frac{1}{r}\frac{\partial}{\partial r}\left(rh^3\frac{\partial p}{\partial r}\right) \qquad (14)$$

where, $r$ is radial coordinate, $t$ is time, $\mu_{air}$ is dynamic viscosity of the air, $p$ is the pressure in the air film and $h(r, t)$ is the distance between the liquid-gas interface and substrate. Using geometry shown in Figure 15(a), we get,

$$h(r,t)=b(t)-\left((d_0/2)^2-r^2\right)^{1/2} \qquad (15)$$

where $b(t)$ is the minimum distance between the liquid-gas interface and substrate. At very close to the impact location, $r/(d_0/2) \ll 1$, eq. 15 is approximated as follows,

$$h(r,t) = b(t) + \frac{d_0}{4}\left(\frac{r^2}{(d_0/2)^2}\right) + O\left(\frac{r^4}{(d_0/2)^4}\right) \qquad (16)$$

Neglecting higher order terms in eq., 16, $h(r, t)$ becomes independent of $r$ i.e. $h(r, t) \sim b(t)$. Integrating eq. 14 with respect to $r$ and using $h(r, t) \sim b(t)$, we obtain,

$$rb^3 \frac{\partial p}{\partial r} = 6\mu_{air} r^2 \frac{\partial b}{\partial t} + c \qquad (17)$$

where, $c$ is a constant of integration. Considering Neumann boundary condition for the pressure at the substrate surface leads to $c = 0$ and eq. 17 is futher simplified as [43],

$$\frac{\partial p}{\partial r} = 6\mu_{air} \frac{r}{h^3} \frac{\partial b}{\partial t} \qquad (18)$$

Since at any instance $t$, $b(t)$ scales as, $b(t) \sim v_0 t$, eq. 18 is rewritten using scaling arguments as shown by Duchemin and Josserand [44], as follows,

$$p_{air} \sim \frac{\mu_{air} r^2 v_0}{b^3} \qquad (19)$$

Considering the deformation of the liquid-gas interface near the impact location due to air overpressure as shown in Figure 15(b), the distance $R(t)$ is expressed using the geometry as follows,

$$(d_0/2)^2 = ((d_0/2) - b)^2 + R^2 \qquad (20)$$

Neglecting higher order terms in eq. 20, $R(t)$ scales as,

$$R \sim \sqrt{d_0 b} \qquad (21)$$

The pressure generated in the squeezed air film at $R$ is given by eq. 19 as follows,

$$p_{air} \sim \frac{\mu_{air} R^2 v_0}{b^3} \qquad (22)$$

Following Duchemin and Josserand [44], and Josserand and Zaleski [45], the droplet impact pressure at $R$ scales with ratio of the rate of change of the momentum of the impacting liquid and wetted area,

$$p_{droplet} \sim \frac{1}{R^2} \frac{d}{dt}\left(\rho R^3 v_0\right) \qquad (23)$$

Eq. 23 is further simplified using eq. 21 and $b(t) \sim v_0 t$, as follows,

$$p_{droplet} \sim \rho v_0^2 \frac{d_0}{R} \tag{24}$$

At the instance of the bubble entrapment, the pressure generated in the squeezed air film (eq. 22) and droplet impact pressure (eq. 24) at $R$ are on the same order i.e $p_{air} \sim p_{droplet}$. Defining this location as $R^*$ and using eqs. 21, 22 and 24, we obtain [44],

$$R^* \sim d_0 \left( \frac{\mu_{air}}{\rho v_0 d_0} \right)^{1/3} \tag{25}$$

Similarly, defining $b^*$ at this instance, we get [44],

$$b^* \sim d_0 \left( \frac{\mu_{air}}{\rho v_0 d_0} \right)^{2/3} \tag{26}$$

Using parameters in the present study, $\mu_{air} = 1.846 \times 10^{-5}$ Pa-s, $v_0 = 0.44$ m/s, $d_0 = 2$ mm, $\rho = 1000$ kg/m$^3$, we get, $R^* = 55$ μm and $b^* = 1.5$ μm. Thus, the maximum deformation of the liquid-gas interface along the axis of symmetry is 1.5 μm. Assuming the shape of entrapped bubble approximately as a spherical cap, the calculated bubble volume is $7.1 \times 10^{-6}$ microliter, which is around six orders of magnitude lesser than the droplet volume.

## 6.2 Mathematical expressions of the energies

The droplet possesses kinetic ($E_k$), surface ($E_s$) and gravitational ($E_g$) energies. The expression of kinetic energy is given by [64, 65]

$$E_k = \frac{1}{2} \rho \int_\Omega \left( u^2 + v^2 \right) d\Omega, \tag{27}$$

where $\rho$ is the droplet density, $u$ and $v$ are the radial and axial components of the velocity, respectively, and $\Omega$ is the droplet volume. The surface energy is given by [14, 64, 65]

$$E_s = \gamma_{LG} A_{LG} + \gamma_{SL} A_{SL} - \gamma_{SG} A_{SL}, \tag{28}$$

where $\gamma_{SG}$, $\gamma_{LS}$ and $\gamma_{LG}$ are the surface tension between the solid-gas, liquid-solid, and liquid-gas, respectively. $A_{SL}$ and $A_{LG}$ are the interfacial areas for the solid-liquid and liquid-gas, respectively. The Young's equation relates the contact angle $\theta$ and the surface energies as follows [73],

$$\gamma_{LG} \cos \theta = \gamma_{SG} - \gamma_{LS}, \tag{29}$$

The expression of surface energy is simplified using eq. 28 and 29, as follows,

$$E_s = \gamma_{LG}\left[ A_{LG} - \pi r_{wetted}^2 \cos\theta \right], \tag{30}$$

where $r_{wetted}$ is the droplet wetted radius. The gravitational energy is given by [64, 65],

$$E_g = \rho g \int_\Omega z\, d\Omega, \tag{31}$$

where $z$ is the axial coordinate of the elemental volume dΩ and $g$ is the acceleration due to gravity.

# 8 Tables

Table 1. Parameters used in the simulation cases.

| Cases | Droplet diameter $d_0$ (mm) | Impact velocity $v_0$ (m s$^{-1}$) | Measurements reported by | Reynolds number Re | Weber number We | Equilibrium contact angle $\theta_{eq}$ (degrees) | Fate of droplet |
|---|---|---|---|---|---|---|---|
| 1 | 2.0 | 0.29 | Tsai et al. [32] | 645 | 2.34 | 155º ± 3º | Bouncing |
| 2 | 2.0 | 0.13 | Tsai et al. [32] | 289 | 0.45 | 155º ± 3º | Non-bouncing |
| 3 | 2.0 | 0.56 | Wang et al. [37] | 1244 | 8.71 | 163º | Bouncing |
| 4 | 2.0 | 0.44 | Jung and Bhushan [31] | 978 | 5.37 | 158º ± 2.4º | Bouncing |
| 5 | 2.0 | 0.44 | Jung and Bhushan [31] | 978 | 5.37 | 91º ± 2º | Non-bouncing |
| 6 | 2.0 | 0.56 | Wang et al. [37] | 1244 | 8.71 | 140º | Non-bouncing* |

* Wetting transition from Cassie to Wenzel state

Table 2: Simulations performed for different Reynolds and Weber numbers at equilibrium contact angle of 155°. The corresponding values of the initial droplet diameter and impact velocity are given for each entry in the table.

| Re→<br>We↓ | 100 | 200 | 300 | 400 | 500 | 600 | 700 | 800 | 900 | 1000 | 1100 | 1200 |
|---|---|---|---|---|---|---|---|---|---|---|---|---|
| 1 | 0.11 mm<br>0.8 m s$^{-1}$ | 0.45<br>0.4 | 1.012<br>0.27 | 1.8<br>0.2 | 2.81<br>0.16 | 4.05<br>0.13 | 5.51<br>0.11 | 7.2<br>0.1 | 9.11<br>0.089 | 11.25<br>0.08 | 13.62<br>0.073 | 16.19<br>0.067 |
| 2 | 0.056<br>1.6 | 0.23<br>0.8 | 0.51<br>0.53 | 0.9<br>0.4 | 1.41<br>0.32 | 2.025<br>0.27 | 2.76<br>0.23 | 3.6<br>0.2 | 4.56<br>0.18 | 5.63<br>0.16 | 6.80<br>0.15 | 8.10<br>0.13 |
| 3 | 0.038<br>2.4 | 0.15<br>1.2 | 0.34<br>0.8 | 0.6<br>0.6 | 0.93<br>0.48 | 1.35<br>0.4 | 1.84<br>0.34 | 2.4<br>0.3 | 3.037<br>0.27 | 3.75<br>0.24 | 4.54<br>0.22 | 5.4<br>0.2 |
| 4 | 0.028<br>3.2 | 0.11<br>1.6 | 0.25<br>1.067 | 0.45<br>0.8 | 0.70<br>0.64 | 1.013<br>0.53 | 1.38<br>0.46 | 1.8<br>0.4 | 2.28<br>0.3556 | 2.81<br>0.32 | 3.40<br>0.29 | 4.045<br>0.27 |
| 5 | 0.023<br>4 | 0.09<br>2 | 0.2<br>1.33 | 0.36<br>1 | 0.56<br>0.8 | 0.81<br>0.67 | 1.10<br>0.57 | 1.44<br>0.5 | 1.82<br>0.44 | 2.25<br>0.4 | 2.72<br>0.36 | 3.24<br>0.33 |
| 6 | 0.019<br>4.8 | 0.075<br>2.4 | 0.17<br>1.6 | 0.3<br>1.2 | 0.47<br>0.96 | 0.68<br>0.8 | 0.92<br>0.69 | 1.2<br>0.6 | 1.52<br>0.53 | 1.88<br>0.48 | 2.27<br>0.44 | 2.7<br>0.4 |
| 7 | 0.016<br>5.6 | 0.064<br>2.8 | 0.15<br>1.87 | 0.26<br>1.4 | 0.40<br>1.12 | 0.58<br>0.93 | 0.79<br>0.8 | 1.029<br>0.7 | 1.30<br>0.62 | 1.61<br>0.56 | 1.95<br>0.51 | 2.31<br>0.47 |

# 9 Figures

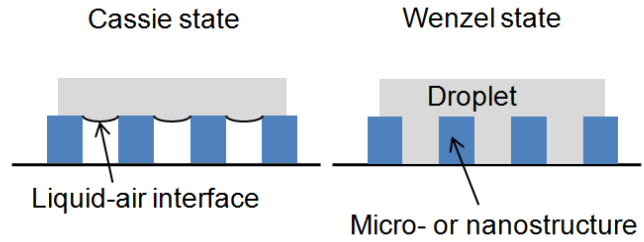

Figure 1: Schematic of Cassie and Wenzel states during droplet impact on a micro- or nanostructured superhydrophobic surface.

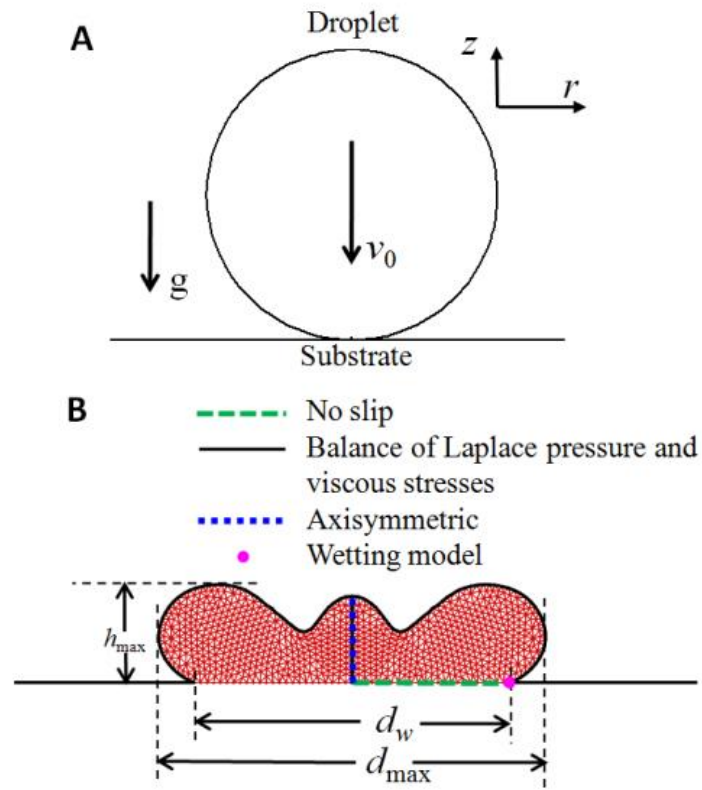

Figure 2: (A) Initial condition of the simulation (B) Boundary conditions and finite-element mesh used in the computational model.

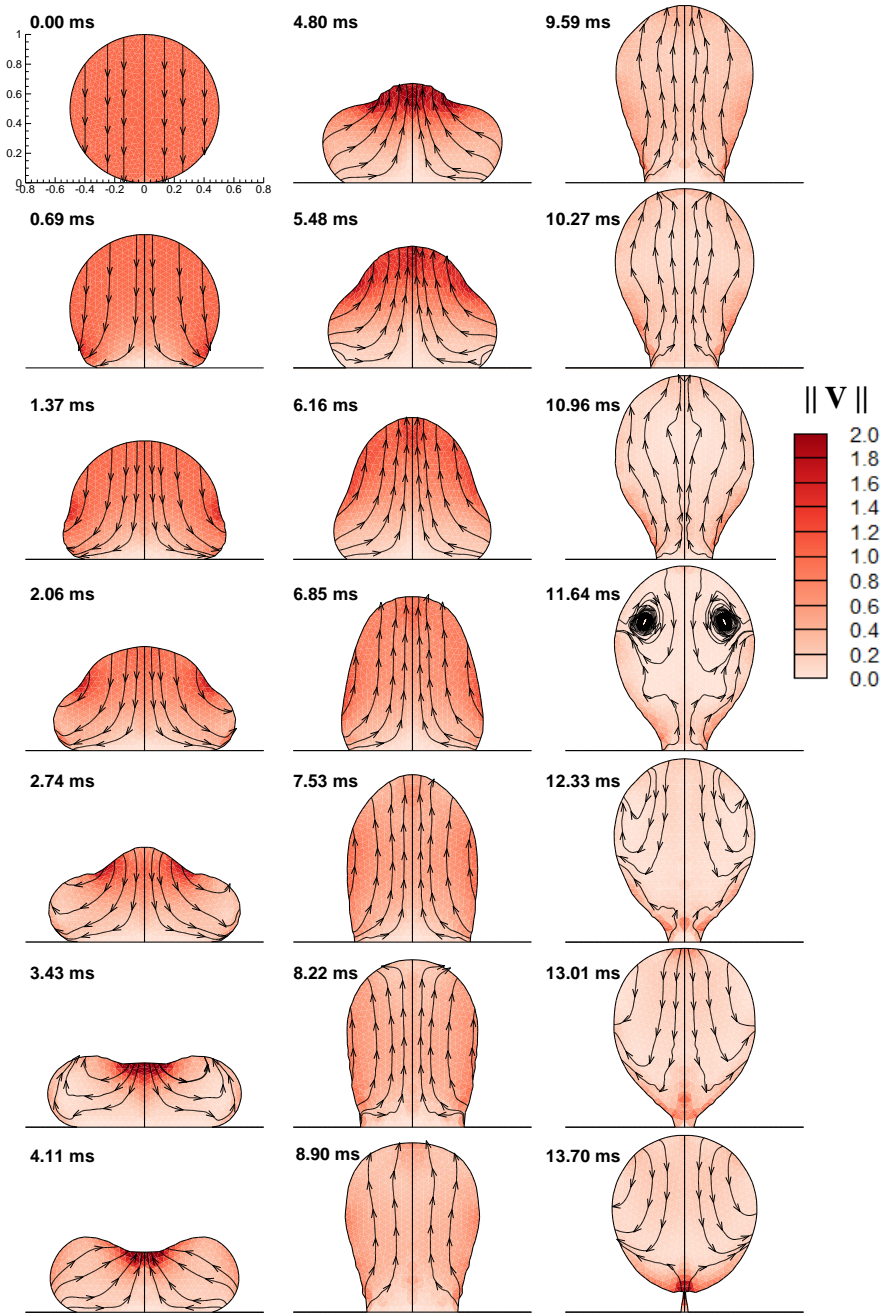

Figure 3: Numerical simulation of bouncing of 2 mm water droplet on superhydrophobic surface with impact velocity and equilibrium contact angle as 0.29 m s$^{-1}$ and 155º, respectively (case 1 in Table 1, $Re$ = 645, $We$ = 2.34). Droplet shapes and instantaneous streamlines are shown at different time instances (see also associated computer animation [62]).

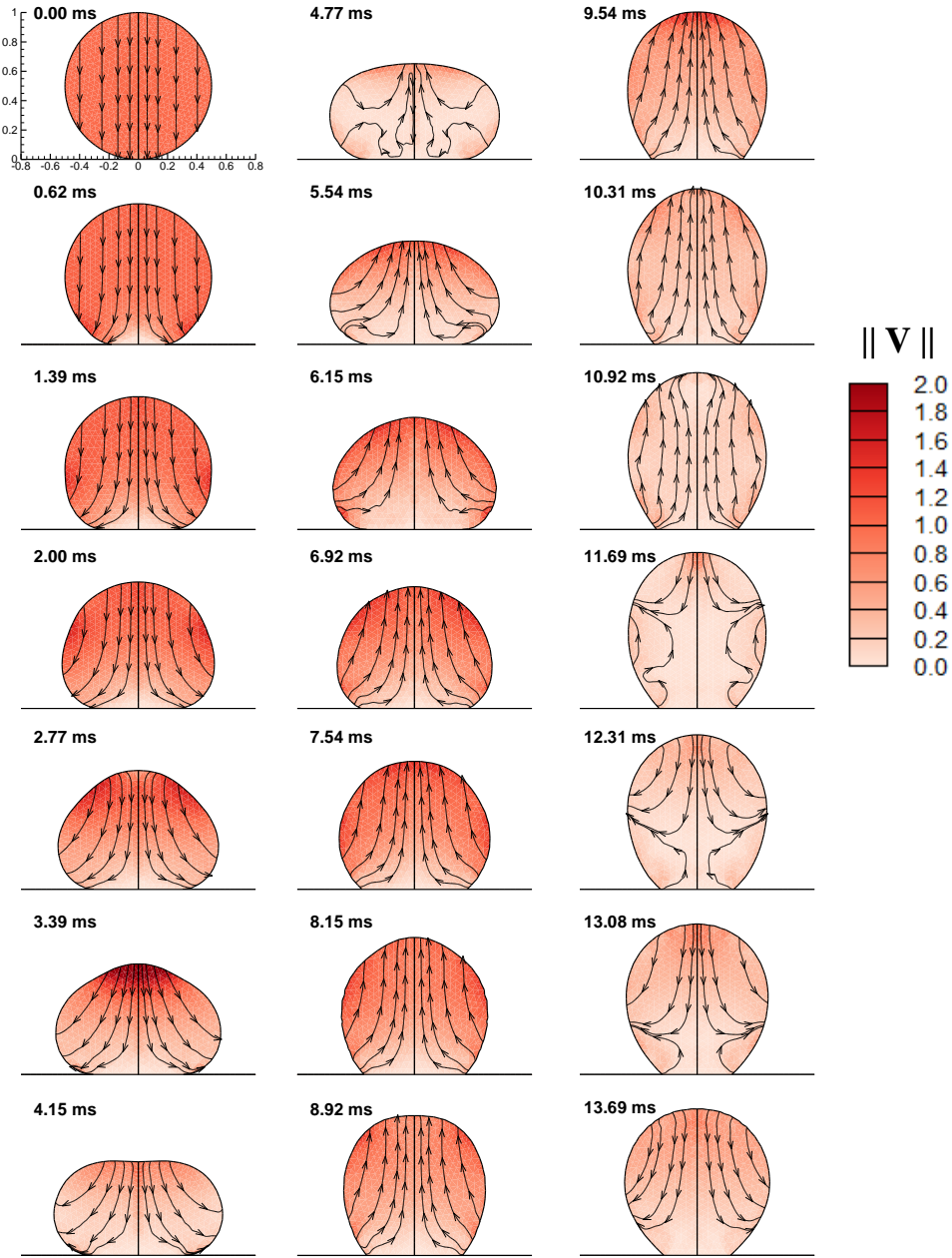

Figure 4 : Numerical simulation of non-bouncing of 2 mm water droplet on superhydrophobic surface with impact velocity and equilibrium contact angle as 0.13 m s$^{-1}$ and 155º, respectively (case 2 in Table 1, *Re* = 289, *We* = 0.45). Droplet shapes and instantaneous streamlines are shown at different time instances (see also associated computer animation [63]).

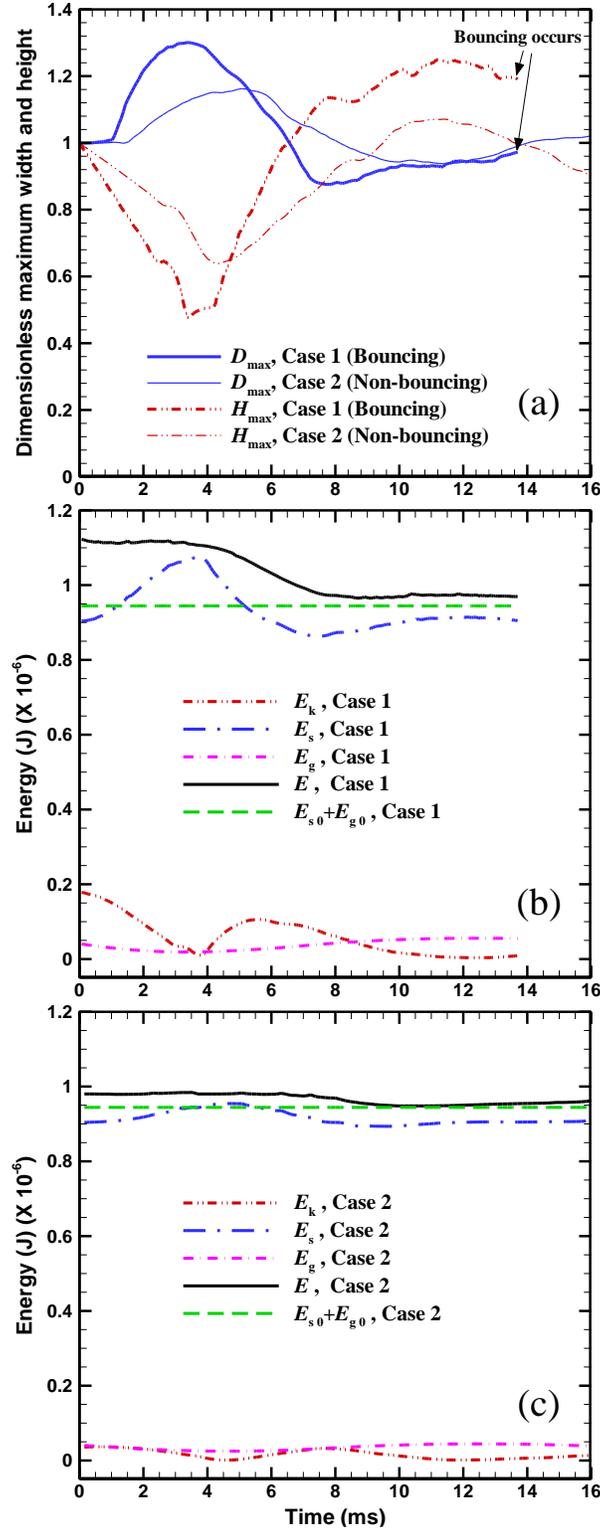

Figure 5: (a) Comparison between time-varying droplet maximum width ($D_{max}$) and height ($H_{max}$) obtained by the numerical simulations for two cases of impact velocity. Cases 1 and 2 correspond to impact velocity of 0.29 and 0.13 m s$^{-1}$, respectively. (b) Time-variation of the kinetic ($E_k$), surface ($E_s$), gravitational ($E_g$) and total ($E = E_k + E_s + E_g$) energy for case 1. The sum of the initial surface and gravitational energy ($E_{s0} + E_{g0}$) is also shown as a horizontal line. (c) Time-variation of the energies as plotted in (b) for case 2.

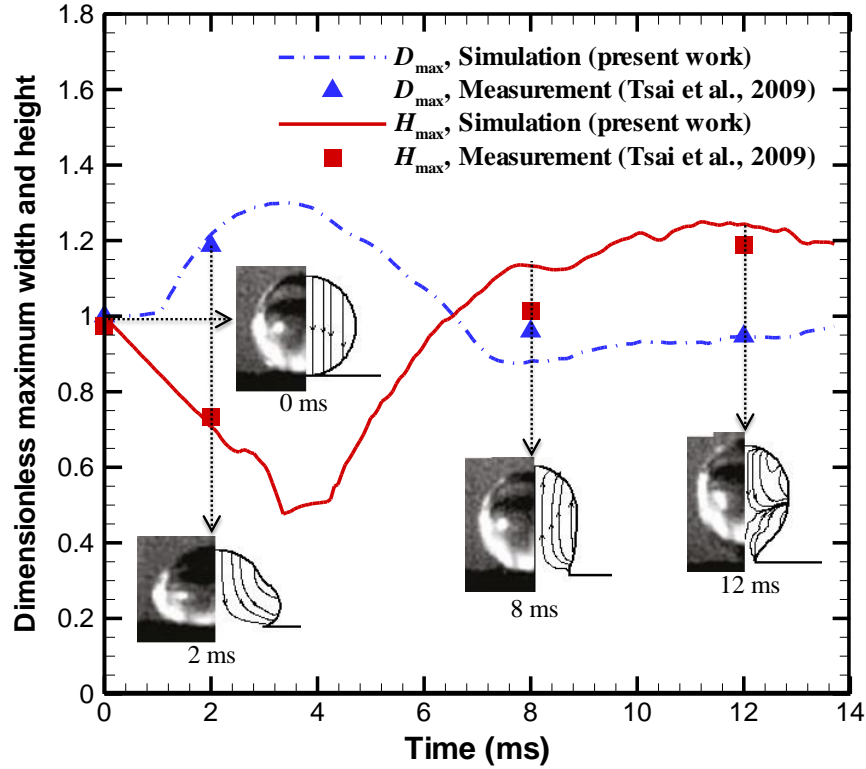

Figure 6: Comparison between time-varying droplet maximum width and height obtained by the numerical simulation and measurement by Tsai et al. [32] for bouncing of 2 mm water droplet (case 1 in Table 1, $Re = 645$, $We = 2.34$, $v_0 = 0.29$ m s$^{-1}$, $\theta_{eq} = 155°$). Lines and symbols represent the simulation and measurement, respectively. Insets show the comparison between droplet shapes obtained in measurements [32] (left) and simulated droplet shapes with instantaneous streamlines (right) at different time instances. Images in insets (left) adapted with permission from [32]. Copyright (2009) American Chemical Society.

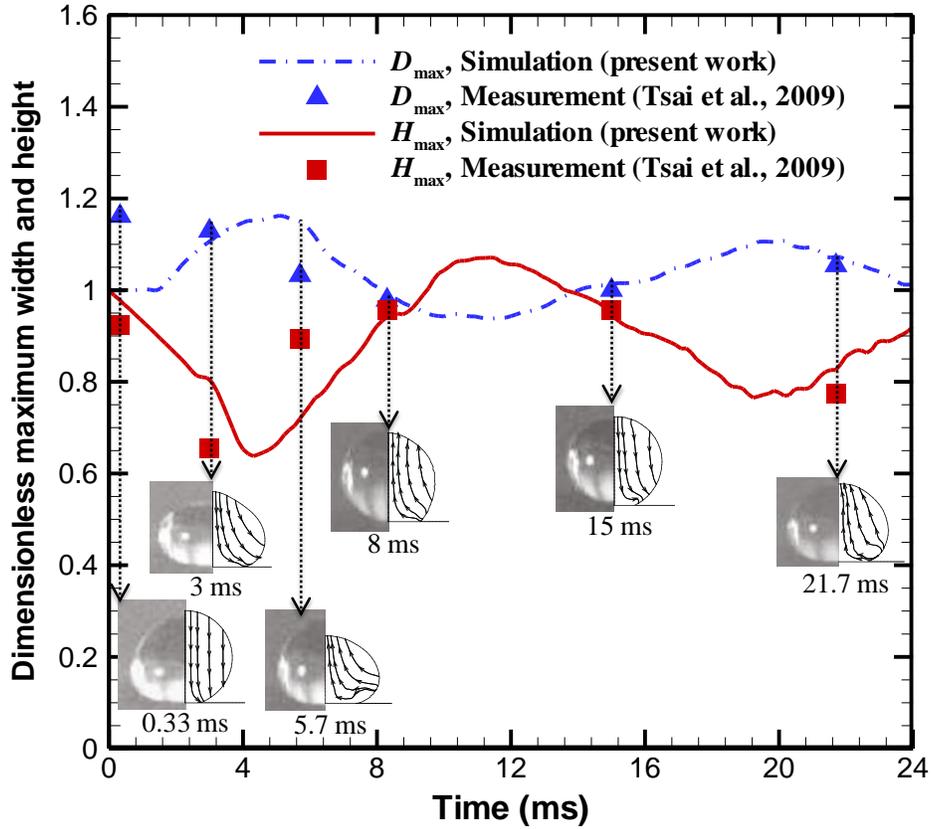

Figure 7 : Comparison between time-varying droplet maximum width and height obtained by the numerical simulation and measurement by Tsai et al. [32] for non-bouncing of 2 mm water droplet (case 2 in Table 1, $Re = 289$, $We = 0.45$, $v_0 = 0.13$ m s$^{-1}$, $\theta_{eq} = 155°$). Lines and symbols represent simulation and measurement, respectively. Insets show the comparison between droplet shapes obtained in measurements [32] (left) and simulated droplet shapes with instantaneous streamlines (right) at different time instances. Images in insets (left) adapted with permission from [32]. Copyright (2009) American Chemical Society.

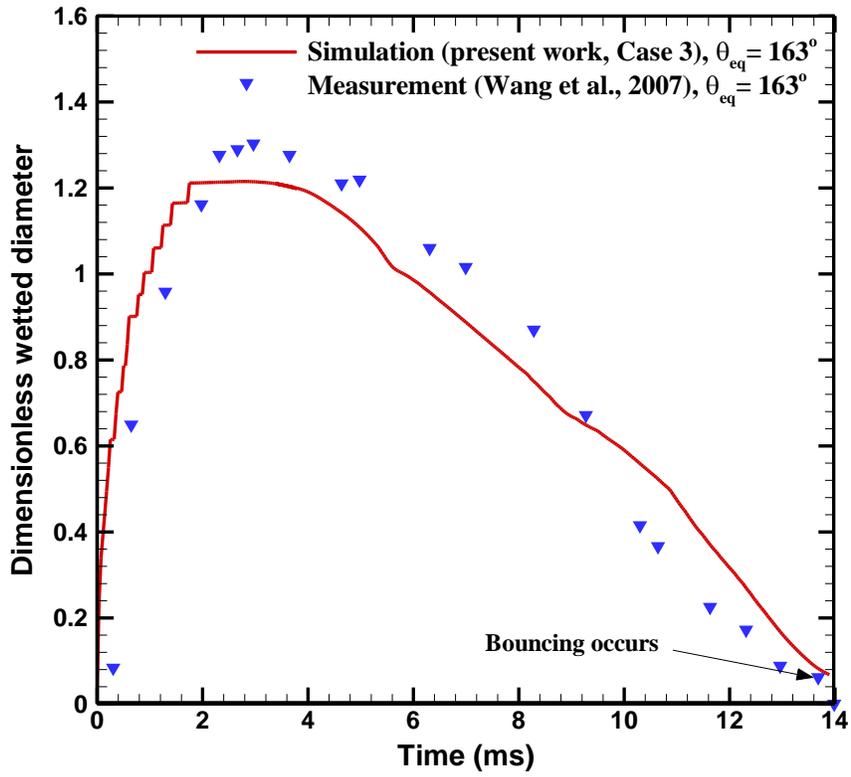

Figure 8: Comparison between time-varying droplet wetted diameter obtained by numerical simulation and measurement by Wang et al. [37] for equilibrium contact angle of 163º (cases 3 in Table 1, $Re = 1244$, $We = 8.71$). Lines and symbols represent the simulation and measurement, respectively.

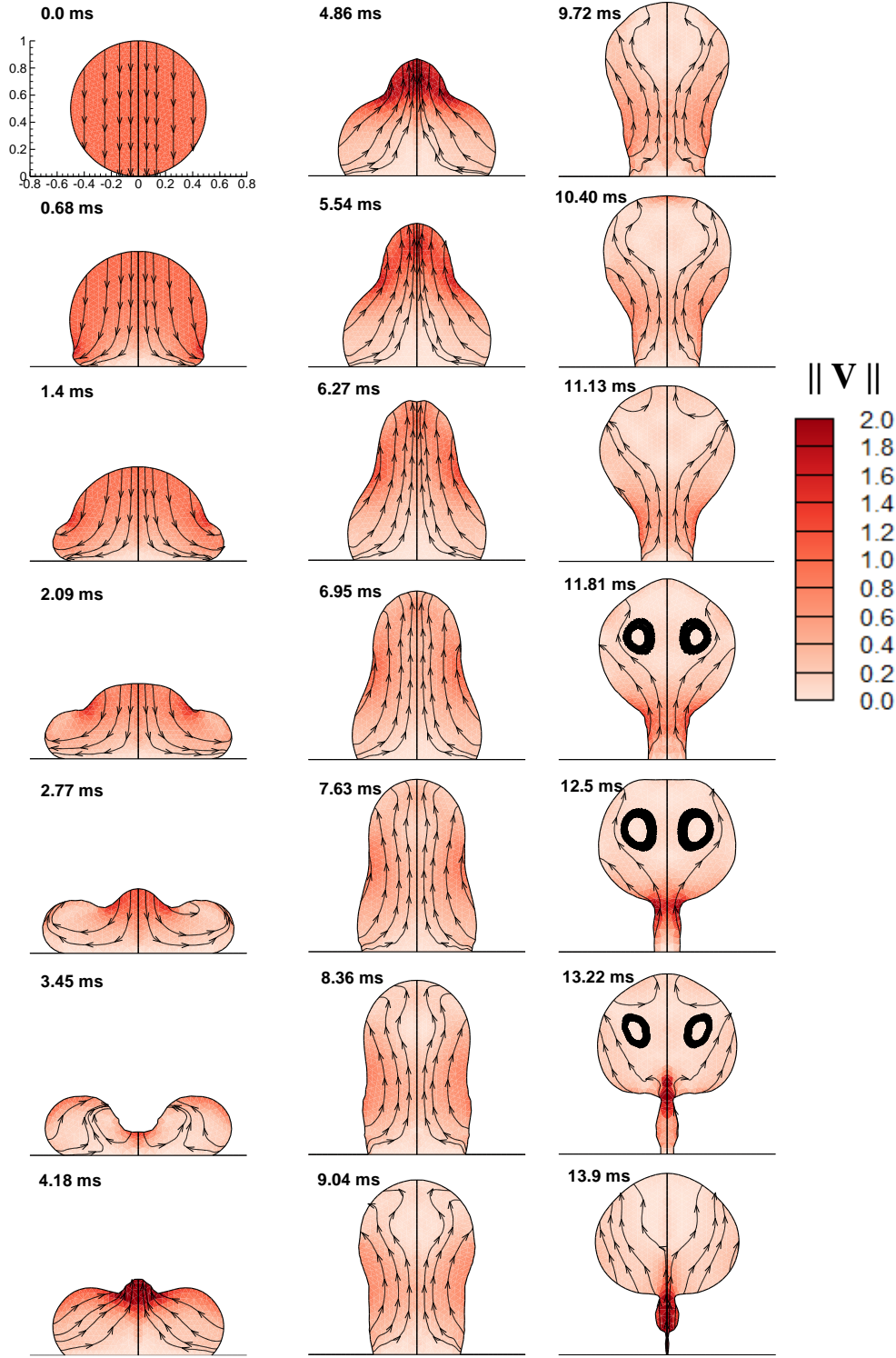

Figure 9 : Numerical simulation of bouncing of 2 mm water droplet on superhydrophobic surface with impact velocity and equilibrium contact angle 0.44 m s$^{-1}$ and 158°, respectively (case 4 in Table 1, *Re* = 978, *We* = 5.37). Droplet shapes and instantaneous streamlines are shown at different instances (see also associated computer animation [67]).

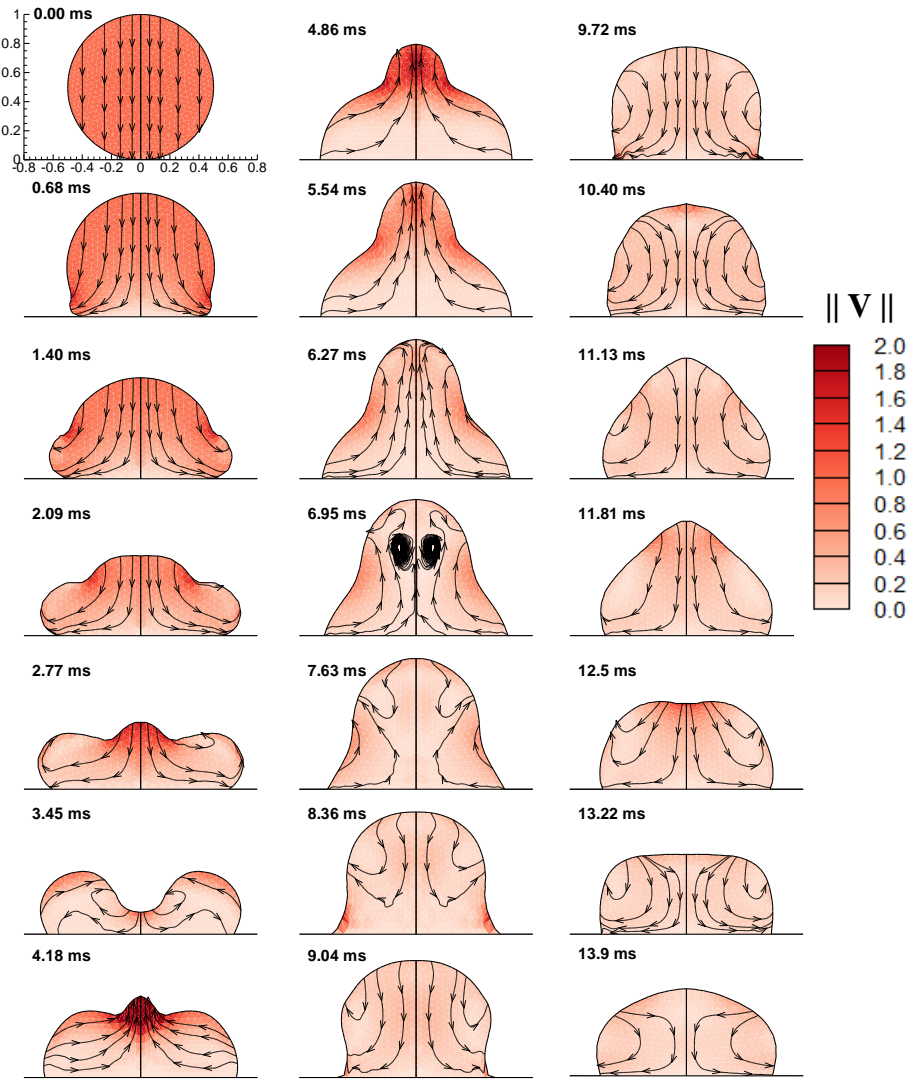

Figure 10 : Numerical simulation of non-bouncing of 2 mm water droplet on superhydrophobic surface with impact velocity and equilibrium contact angle as 0.44 m s$^{-1}$ and 91º, respectively (case 5 in Table 1, *Re* = 978, *We* = 5.37). Droplet shapes and instantaneous streamlines are shown at different instances (see also associated computer animation [68]).

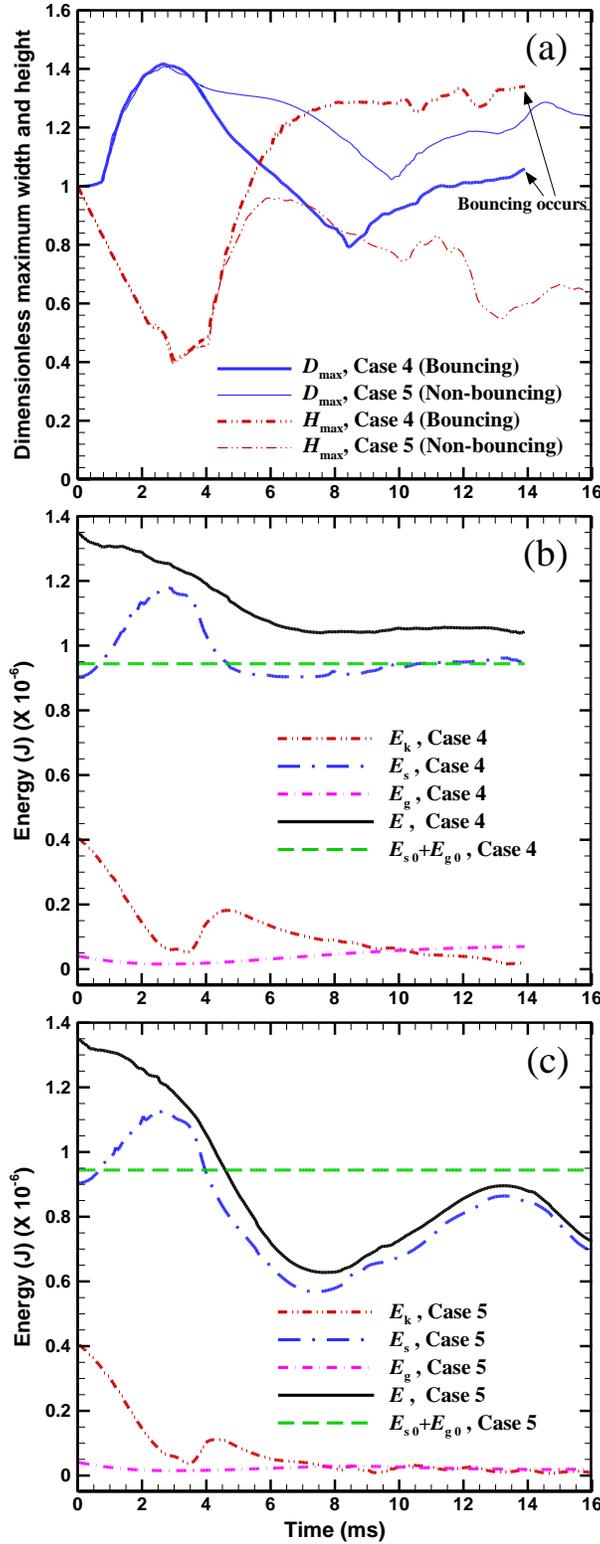

Figure 11: (a) Comparison between time-varying droplet maximum width and height obtained by the numerical simulations for two cases of equilibrium contact angle. Cases 4 and 5 correspond to equilibrium contact angle of 158º and 91º, respectively. (b) Time-variation of the kinetic ($E_k$),

surface ($E_s$), gravitational ($E_g$) and total ($E = E_k + E_s + E_g$) energy for case 4. The sum of the initial surface and gravitational energy ($E_{s0} + E_{g0}$) is also shown as a horizontal line. (c) Time-variation of the energies as plotted in (b) for case 5.

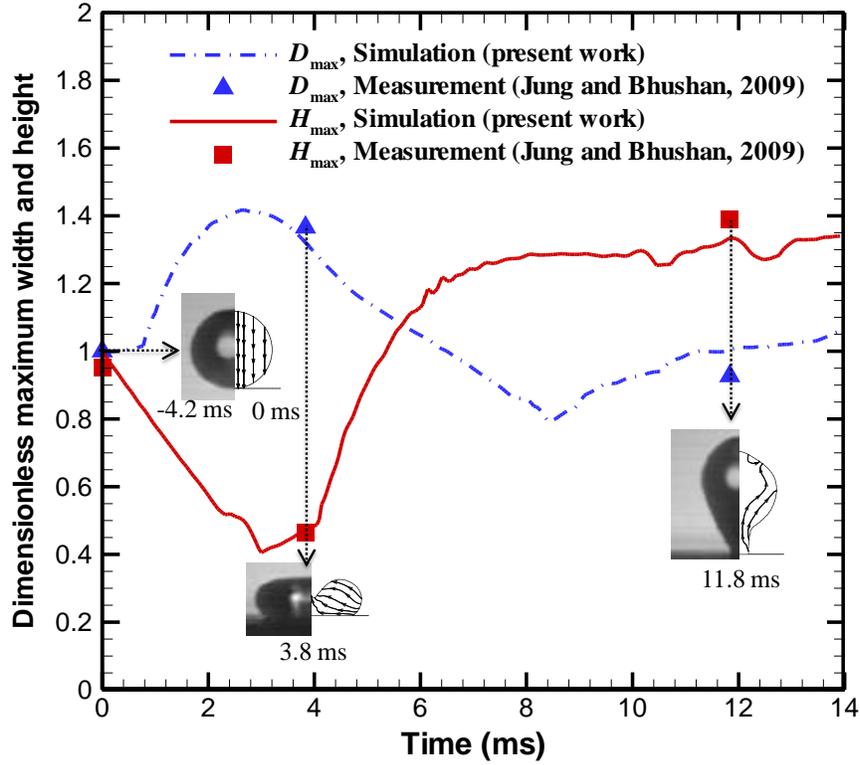

Figure 12 : Comparison between time-varying droplet maximum width and height, obtained by the numerical simulation and measurement by Jung and Bhushan [31] for bouncing of 2 mm water droplet (case 4 in Table 1, $Re = 978$, $We = 5.37$, $v_0 = 0.44$ m s$^{-1}$, $\theta_{eq} = 158°$). Lines and symbols represent simulation and measurement, respectively. Insets show the comparison between droplet shapes at different instances obtained in measurement [31] (left) and simulated droplet shapes with instantaneous streamlines (right). Images in insets (left) adapted with permission from [31]. Copyright (2009) American Chemical Society.

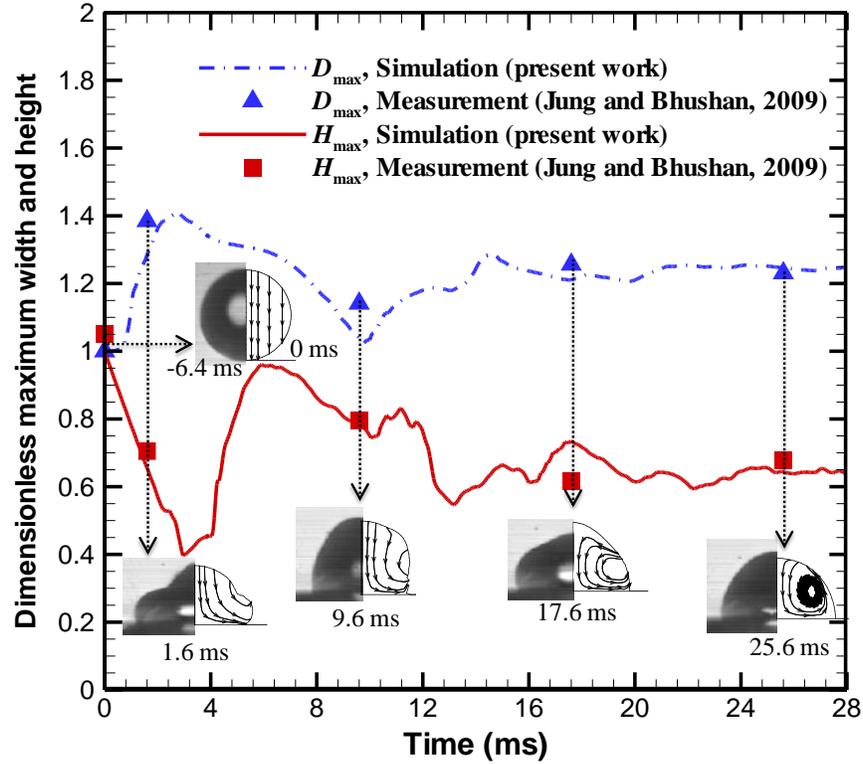

Figure 13: Comparison between time-varying droplet maximum width and height obtained by the numerical simulation and measurement by Jung and Bhushan [31] for non-bouncing of 2 mm water droplet (case 5 in Table 1, $Re = 978$, $We = 5.37$, $v_0 = 0.44$ m s$^{-1}$, $\theta_{eq} = 91°$). Lines and symbols represent simulation and measurement, respectively. Insets show the comparison between droplet shapes at different instances obtained in measurement [31] (left) and simulated droplet shapes with instantaneous streamlines (right). Images in insets (left) adapted with permission from [31]. Copyright (2009) American Chemical Society.

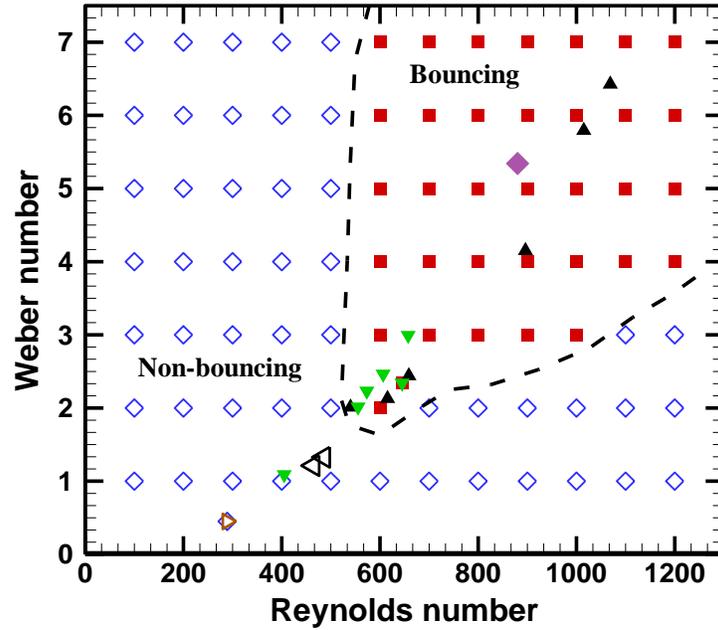

Figure 14: Regime map to predict the bouncing and non-bouncing droplets on a superhydrophobic surface with equilibrium contact angle of around 155º. The bouncing and non-bouncing cases are represented by filled and hollow symbols, respectively, and the numerical as well as available measurements are plotted. The measurements correspond to equilibrium contact angle of 155º± 3º and dashed line is plotted to demarcate the regimes.

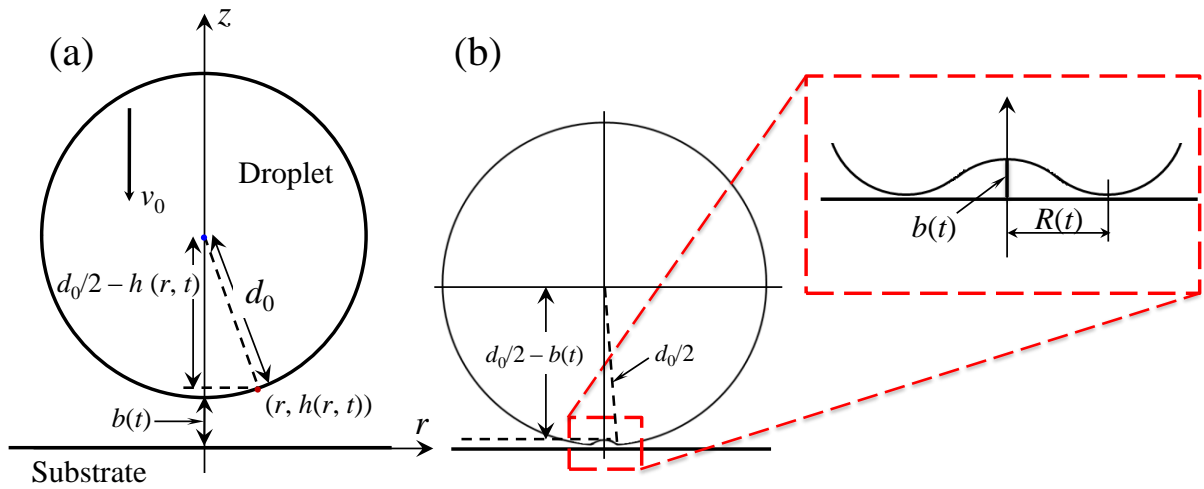

Figure 15: Schematic for analytical model for calculating liquid-gas deformation due to air squeezing and size of entrapped bubble (a) Initial condition of the droplet when air effects become important (b) The possible shape of the liquid-gas interface very close to the substrate.

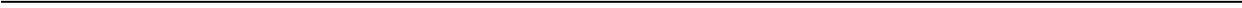